\documentclass[12pt]{amsart}
\usepackage{lipsum}
\usepackage{amsfonts}
\usepackage{graphicx}
\usepackage{hyperref,bm,comment} 
\usepackage{subcaption}
\usepackage{epstopdf}
\usepackage{amssymb,latexsym}
\usepackage{cite}
\usepackage{cases}
\usepackage[utf8]{inputenc}
\usepackage{calrsfs}
\usepackage{algorithm} 
\usepackage{algpseudocode}
\usepackage{booktabs} 
\usepackage{siunitx} 
\usepackage{caption} 
\usepackage{geometry} 
\usepackage{mathrsfs}  
\usepackage[T1]{fontenc}
\usepackage{lmodern}
\usepackage{amsmath,amssymb,booktabs,graphicx}
\usepackage{float}
\hypersetup{colorlinks=true,linkcolor=blue,citecolor=blue,urlcolor=blue}
\usepackage{microtype}
\DeclareMathAlphabet{\pazocal}{OMS}{zplm}{m}{n}
\ifpdf
\DeclareGraphicsExtensions{.eps,.pdf,.png,.jpg}
\else
\DeclareGraphicsExtensions{.eps}
\fi

\usepackage{enumitem}
\setlist[enumerate]{leftmargin=.5in}
\setlist[itemize]{leftmargin=.5in}

\newcommand{\mb}{\begin{bmatrix}}
	\newcommand{\me}{\end{bmatrix}}
\newtheorem{thm}{Theorem}[section]

\newtheorem{prop}[thm]{Proposition}

\DeclareMathOperator{\E}{\mathbb{E}}
\DeclareSymbolFont{largesymbol}{OMX}{yhex}{m}{n}
\DeclareMathAccent{\Widehat}{\mathord}{largesymbol}{"62}
\numberwithin{equation}{section}

\title[Gated payoff-weighted net profit maximization]{Gated payoff-weighted net profit maximization for online portfolio selection with transaction costs}
\author{Grace Chengrong Xie}
\address{Department of Mathematics, University of Hong Kong, Pokfulam, Hong Kong}
\email{gracexiecr@connect.hku.hk}

\author{Alvey Qianli Lin}
\address{Department of Mathematics, University of Hong Kong, Pokfulam, Hong Kong}
\email{alveyqlin@connect.hku.hk}

\author{Wai Ki Ching}
\address{Department of Mathematics, University of Hong Kong, Pokfulam, Hong Kong}
\email{wching@hku.hk}

\date{\today}

\usepackage{amsopn}

\usepackage{xcolor}

\ifpdf
\hypersetup{
	pdftitle={Gated payoff-weighted net profit maximization for online portfolio selection with transaction costs},
	pdfauthor={A. Q. Lin}
}
\fi

\begin{document}

\maketitle

\begin{abstract}
In this paper, we study online portfolio selection (OLPS) problems. 
The net profit maximization (NPM) acting upon Adaptive Online Moving Average (AOLMA), namely the AOLNPM strategy in portfolio selection, trades when forecast gain exceeds the transaction cost. In essence, this method relies heavily on calibrated forecasts. Indeed, we show that AOLMA forecasts overstate the realized relative returns. 
We therefore design a gate that adjusts how strongly an AOLMA forecast is used; for each asset, it learns the recent mixture proportion assigned to AOLMA from predictive likelihoods against a neutral forecast (no change). The proportion therefore estimates the extent to which the forecast is realized. We then feed the resulting forecast into the NPM portfolio rule, where the gate acts as an asset-specific cost threshold. A payoff-weighted fund combines gated NPM experts with fixed decay factors in proportion to their wealth and needs no learning rate. On nine datasets with a transaction cost rate of \(0.1\%\), our method produces \(5.9\) times the wealth of AOLNPM in geometric mean and sits above AOLNPM on every dataset, with its preponderance growing with the cost rate.
\end{abstract}

\section{Introduction}\label{sec:intro}

Online portfolio selection (OLPS) is an important and attractive research problem for both academic researchers and practitioners. 
In classical portfolio selection theory \cite{markowitz1952portfolio}, it is assumed that the risky asset return is subject to a certain distribution function. 
Under the assumption of distribution function, one can
calculate expected value and variance of the return 
and therefore the expected return and risk. 
The investors then allocate the capital in different assets to achieve better investment return or avert the risk.
OLPS is different from the classical portfolio selection theory
in that it does not assume any return distribution and it is also a data-driven
sequence decision process. 
Using only past prices, OLPS regularly adjusts a portfolio of assets. There are many different strategies available in the literature, see for instance, \cite{guo2026data,li2014online,li2018online}. 

The Adaptive Online Net Profit Maximization (AOLNPM) \cite{guo2021adaptive} addresses the transaction cost directly and is further extended with state-dependent price estimators \cite{guo2023online}. 
The method applies an Adaptive Online Moving Average (AOLMA) to forecast each price relative and chooses the portfolio by NPM, which maximizes the forecast return net of the transaction cost. 
The solution of NPM has a simple form and a straightforward standard, which is to sell every holding whose forecast lags the best forecast by more than a specified cost rate \(c\) and relocate the portfolio to the best asset (Proposition~\ref{prop:npm}).
However, this rule takes the forecasts at face value. 
A switch from one asset to another raises expected net return only if the expected return gap exceeds the transaction cost; otherwise it would lead to a greater number of unnecessary trading changes. If only a fraction \(\beta\) of the forecast gap is realized, the threshold should be adjusted accordingly to \(c/\beta\) (Proposition~\ref{prop:calib}). Our test shows that AOLMA indeed falls into the limitations. On the four benchmark datasets of \cite{guo2021adaptive}, the asset with the highest forecast only realizes \(14\)--\(30\%\) of its forecast excess over the average asset. 
On five additional datasets, the ratio is at most \(7\%\) (Table~\ref{tab:calib}). 
Therefore, this leads to NPM trading on gaps which do not cover the cost. 
Though AOLNPM adapts its decay factors to the forecast error, the step sizes turn out to be small and data-specific. 
The final decay factors in our runs lie between \(0.51\) and \(0.72\).

From the above observations, in this paper, 
we propose two adjustments for improvement. 
The first is a gate which adjusts how strongly to adopt the AOLMA forecast. 
The method learns the recent mixture proportion assigned to the AOLMA forecast from predictive likelihoods against a neutral forecast for each asset. 
Then the resulting forecast proceeds to enter the NPM portfolio rule. The gate calibrates each forecast online by using the proportion to estimate the share of the forecast that is realized (Proposition~\ref{prop:pibeta}). 
Within NPM, the gate then serves as an asset-specific threshold and re-ranks the assets by the reliability of their signals (Proposition~\ref{prop:gatenpm}). 
The second adjustment is to replace the adaptive decay factor by a 
payoff-weighted fund. 
We run gated NPM experts with fixed decay factors on a grid and weight them in proportion to the wealth they have earned. 
Hence, no learning rate is required for the fund. 
In separate accounts, at least a fixed fraction of the wealth of the best expert is retained (Proposition~\ref{prop:fund}). 
We call this method the Gated Payoff-weighted NPM (G-PWNPM). 

Our numerical experiments are run on nine datasets 
with transaction cost rate \(c=0.001\). 
On geometric mean, G-PWNPM ends with \(5.92\) times the terminal wealth of AOLNPM and sits above it on all nine datasets, with pooled statistic \(z=4.68\). On the four benchmark datasets among them, G-PWNPM sits above all thirteen baselines. Furthermore, on the five additional datasets where forecasts carry little information, it improves on AOLNPM but sits in the middle of the baselines. 
One finding is that the preponderance of our method over AOLNPM grows with the cost rate, and that our method has lower average one-way turnover. The first change, the gate, supplies most of the gain mainly by re-ranking the assets, and the fund supplies consistency across datasets.

The paper is structured as follows: 
Section~\ref{sec:prelim} reviews the framework AOLMA, NPM and AOLNPM and states the calibration conditions. 
Section~\ref{sec:models} introduces the two forecast models and their likelihoods.
Section~\ref{sec:mixture} introduces the adaptive mixture weights and Section~\ref{sec:fund} discusses the payoff-weighted fund. 
Numerical experiments are reported in Section~\ref{sec:results}
and Section~\ref{sec:conclusion} concludes the paper.

\section{Preliminaries}\label{sec:prelim}

\subsection{Online portfolio selection with transaction costs}

Let \(n\) be the number of assets and \(P_{ti}\) the price of asset \(i\)
at the end of period \(t\). Its price relative is
\(r_{ti}=P_{ti}/P_{t-1,i}>0\).
Let \(\mathbf b_t\) be the chosen asset-weight vector,
\(\mathbf h_t\) the pre-trade weights carried from the previous period,
and \(c\) the transaction cost rate.
The feasible set is
$$\Delta_n=\{\mathbf b\geq0:\sum_{i=1}^n b_i=1\}.
$$
The weights \(\mathbf b_t\) are chosen before \(\mathbf r_t=(r_{t1},\ldots,r_{tn})^{\mathsf T}\) is observed, and the wealth evolves 
according to
\begin{equation}
 W_t=W_{t-1}
 \left(\mathbf b_t^{\mathsf T}\mathbf r_t
       -\frac c2\|\mathbf b_t-\mathbf h_t\|_1\right),\qquad W_0=1.
 \label{eq:wealth}
\end{equation}
Here \(W_t\) is wealth after period \(t\), and
\(\|\mathbf b_t-\mathbf h_t\|_1/2\) is one-way turnover.
Each unit of wealth bought or sold costs \(c/2\), so moving one unit of wealth from one asset to another costs \(c\); this is the convention of \cite{guo2021adaptive,li2018online}.
After observing the return, the next pre-trade weight is
\(h_{t+1,i}=b_{ti}r_{ti}/(\mathbf b_t^{\mathsf T}\mathbf r_t)\), and \(\mathbf h_1=\mathbf 1/n\).

\subsection{AOLMA forecasts}

The online moving average predicts the next price of asset \(i\) by an exponential average of its past prices, 
$$
\widehat P_{ti}=\alpha P_{t-1,i}+(1-\alpha)\widehat P_{t-1,i}
$$ 
with the decay factor \(\alpha\in[0,1]\), and its price relative 
given by \(f_{ti}=\widehat P_{ti}/P_{t-1,i}\) \cite{li2015moving}. 
Dividing by \(P_{t-1,i}\) and using 
$$
\widehat P_{t-1,i}/P_{t-1,i}=f_{t-1,i}/r_{t-1,i}
$$
we then obtain
\begin{equation}
 y_{ti}=\frac{f_{t-1,i}}{r_{t-1,i}}
\qquad {\rm where} \qquad 
 f_{ti}=\alpha+(1-\alpha)y_{ti}.
 \label{eq:olma}
\end{equation}
The previous price estimate, expressed relative to the latest known
price, is \(y_{ti}\). Unrolling \eqref{eq:olma}, \(f_{ti}\) is a weighted average of past prices divided by the latest price, so \(f_{ti}>1\) when the price lies below its recent average. 
To first order in the returns,
$$
f_{ti}-1\approx-\sum_{k\geq1}(1-\alpha)^k(r_{t-k,i}-1),
$$ 
a small \(\alpha\) gives a long memory and a large reversal signal, and \(\alpha\to1\) gives the neutral forecast \(1\).

The Adaptive Online Moving Average (AOLMA) proposed in \cite{guo2021adaptive} 
allows each asset to have its own decay factor \(\alpha_{ti}\), starting from \(\alpha_{1i}=1/2\). 
After \(r_{ti}\) is observed,
\begin{equation}
 \alpha_{t+1,i}=\alpha_{ti}+\gamma s_{ti},\qquad
 s_{ti}=\begin{cases}\phantom{-}1,&\text{if }\mathbb I\{r_{ti}>f_{ti}\}=\mathbb I\{r_{t-1,i}>f_{t-1,i}\},\\-1,&\text{otherwise,}\end{cases}
 \label{eq:aolma}
\end{equation}
where \(\mathbb I\{A\}\) is \(1\) if \(A\) holds and \(0\) otherwise, and \(\alpha_{t+1,i}\) is reset to \(1/2\) if it leaves \([0,1]\). 
Thus, \(s_{ti}=1\) if and only if the last two forecast errors are both positive or both non-positive. 
Since \(\partial f_{ti}/\partial\alpha_{ti}=1-y_{ti}\) has the sign of \(r_{t-1,i}-f_{t-1,i}\), \eqref{eq:aolma} is a signed gradient step on the squared forecast error \((r_{ti}-f_{ti})^2\) when both forecast errors are nonzero; these are the four cases of \cite[Section 3.2]{guo2021adaptive}. The step size \(\gamma\) is set for each dataset.

\subsection{Net profit maximization}

With forecasts \(\mathbf f_t=(f_{t1},\ldots,f_{tn})^{\mathsf T}\), the NPM rule is
\begin{equation}
 \mathbf b_t\in\arg\max_{\mathbf b\in\Delta_n}
 \left\{\mathbf b^{\mathsf T}\mathbf f_t
       -\frac c2\|\mathbf b-\mathbf h_t\|_1\right\}.
 \label{eq:npm}
\end{equation}
The adaptive online net profit maximization (AOLNPM) strategy proposed in 
\cite{guo2021adaptive} applies \eqref{eq:npm} to AOLMA forecasts and solves it as a linear program. The program has a closed-form solution.

\begin{prop}\label{prop:npm}
Let \(j^*\in\arg\max_i f_{ti}\) and \(S=\{i\neq j^*: f_{tj^*}-f_{ti}>c\}\). An optimal solution of \eqref{eq:npm} is \(b_{ti}=0\) for \(i\in S\), \(b_{ti}=h_{ti}\) for \(i\notin S\cup\{j^*\}\), and \(b_{tj^*}=h_{tj^*}+\sum_{l\in S}h_{tl}\).
\end{prop}

\begin{proof}
Let \(\mu=f_{tj^*}-c/2\) and
\(L(\mathbf b)=\mu+\sum_i\phi_i(b_i)\) with 
$$
\phi_i(b_i)=b_i(f_{ti}-\mu)-\frac c2|b_i-h_{ti}|.
$$ 
On \(\Delta_n\), \(L\) coincides with the objective of \eqref{eq:npm}. 
Each \(\phi_i\) is concave and piecewise linear, 
with slope \(f_{ti}-f_{tj^*}+c\) for \(b_i<h_{ti}\) 
and \(f_{ti}-f_{tj^*}\) for \(b_i>h_{ti}\). 
Hence, \(\phi_{j^*}\) is maximized by every \(b_{j^*}\geq h_{tj^*}\); for \(i\neq j^*\), \(\phi_i\) is maximized by \(b_i=0\) if \(f_{tj^*}-f_{ti}>c\) and by \(b_i=h_{ti}\) otherwise. 
The vector of the proposition maximizes every \(\phi_i\) over \(b_i\ge0\) and lies in \(\Delta_n\), so it maximizes \(L\) over the non-negative orthant and the objective in Eq. \eqref{eq:npm} over \(\Delta_n\).
\end{proof}

In short, NPM sells every holding whose forecast trails the best asset 
by more than a threshold \(c\) and relocates the portfolio 
more heavily towards the best asset. 
Moving \(\delta\) of wealth from asset \(i\) to asset \(j\) changes the objective by \(\delta(f_{tj}-f_{ti}-c)\), 
so the threshold \(c\) is the cost of the move.

\subsection{Calibration of the NPM threshold}

\begin{prop}\label{prop:calib}
Suppose that, given the past, \(\E[r_{tj}-r_{ti}]=\beta(f_{tj}-f_{ti})\) for some \(\beta>0\). Then moving wealth from asset \(i\) to asset \(j\) raises the expected net return if and only if \(f_{tj}-f_{ti}>c/\beta\).
\end{prop}

\begin{proof}
Moving \(\delta\) of wealth gains \(\beta\delta(f_{tj}-f_{ti})\) in expectation and costs \(\delta c\).
\end{proof}

Therefore, under the proportional calibration assumption in 
Proposition~\ref{prop:calib}, the threshold for the forecast gap is \(c/\beta\), which reduces to the NPM threshold \(c\) when \(\beta=1\). 
Table~\ref{tab:calib} estimates \(\beta\) for AOLMA forecasts with \(\alpha=1/2\): the asset with the highest forecast realizes only a small fraction of its forecast excess over the average asset. 
This can lead NPM to trade on forecast gaps whose expected gains do not cover the cost. 
Two remedies are motivated by Proposition~\ref{prop:calib}: raise the threshold to about \(c/\beta\), or rescale each forecast to \(1+\beta_i(f_{ti}-1)\) with an asset-specific \(\beta_i\) and keep the threshold \(c\). The gate of Sections~\ref{sec:models} and~\ref{sec:mixture} does the second, with \(\beta_i\) learned online.

\begin{table}[htbp]
\centering
\small
\begin{tabular}{lrrr}
\toprule
Dataset & Forecast excess & Realized excess & \(\beta\)\\
\midrule
TSE & $0.0437$ & $0.0064$ & $0.15$ \\
MSCI & $0.0131$ & $0.0033$ & $0.26$ \\
NYSE-N & $0.0226$ & $0.0031$ & $0.14$ \\
NYSE-O & $0.0243$ & $0.0074$ & $0.30$ \\
\addlinespace[2pt]
DJIA & $0.0291$ & $0.0010$ & $0.04$ \\
SP500 & $0.0266$ & $0.0019$ & $0.07$ \\
SZ50 & $0.0153$ & $-0.0009$ & $-0.06$ \\
COVID & $0.0192$ & $-0.0006$ & $-0.03$ \\
new-DJIA & $0.0170$ & $0.0001$ & $0.00$ \\
\bottomrule
\end{tabular}
\caption{Calibration of AOLMA forecasts with \(\alpha=1/2\). Forecast excess: forecast of the asset with the highest forecast minus the cross-sectional mean forecast. Realized excess: its price relative minus the cross-sectional mean. Both are averaged over periods; \(\beta\) is their ratio.}
\label{tab:calib}
\end{table}

\section{Two forecast models and their likelihoods}\label{sec:models}

Before period \(t\), an AOLMA forecaster supplies the background forecast \(f^B_{ti}\) of \(r_{ti}\). 
Superscript \(B\) identifies this background. 
It is the forecast of AOLNPM \cite{guo2021adaptive} in this section and the next, and the forecast of a fixed-decay expert in Section~\ref{sec:fund}.
Its decay factor is \(\alpha^B_{ti}\in[0,1]\).
The previous price estimate, expressed relative to the latest known
price, is \(y^B_{ti}\). The recurrence is
\begin{equation}
 y^B_{ti}=\frac{f^B_{t-1,i}}{r_{t-1,i}},\qquad
 f^B_{ti}=\alpha^B_{ti}+(1-\alpha^B_{ti})y^B_{ti}.
 \label{eq:background}
\end{equation}
We compare two prediction models. Model \(1\) uses \(f^B_{ti}\).
Model \(0\) uses the neutral forecast \(1\), corresponding to zero expected
simple return. 
Let \(w_{ti}\in[0,1]\) be the weight assigned to model \(1\)
before trading. The combined forecast is
\begin{equation}
 \widehat f_{ti}
 =w_{ti}f^B_{ti}+(1-w_{ti})\cdot1
 =1+w_{ti}(f^B_{ti}-1).
 \label{eq:forecast}
\end{equation}
Thus, \(w_{ti}=1\) keeps the AOLMA forecast and \(w_{ti}=0\) gives the
neutral forecast. The combined forecasts enter the NPM rule \eqref{eq:npm}.

To assess each model using observed returns, let the hidden label
\(Z_{ti}\in\{0,1\}\) identify its component distribution.
Write \(f_{1ti}=f^B_{ti}\) and \(f_{0ti}=1\).
Let \(v_{ti}>0\) be a shared variance proxy for the log return.
Its square root describes the one-period noise scale.
The conditional working model is
\begin{equation}
 \log r_{ti}\mid Z_{ti}=k
 \sim N\!\left(\log f_{kti}-\frac{v_{ti}}2,v_{ti}\right),
 \qquad k\in\{0,1\},
 \label{eq:emission}
\end{equation}
with mean \(\log f_{kti}-v_{ti}/2\) and variance \(v_{ti}\).
The correction \(-v_{ti}/2\) makes the mean price relative equal to
\(f_{kti}\). We estimate the recent second moment of the log return by
exponential averaging:
\begin{equation}
 v_{t+1,i}=\rho_v v_{ti}+(1-\rho_v)(\log r_{ti})^2,\qquad \rho_v=0.98.
\label{eq:variance}
\end{equation}
Each update combines the previous estimate with the newly observed 
squared log return. 
Repeated updates give exponentially decreasing weights to earlier observations. 
We adopt this second moment as a common variance proxy in \eqref{eq:emission}. Here the identity \(E[X^2]=\operatorname{Var}(X)+(E[X])^2\) 
explains the approximation:
the second moment approaches the variance when the local mean is small.
The first six observations initialize \(v\), with a numerical floor
\(10^{-12}\).

For model \(k\), the log return has a Gaussian density specified by
\eqref{eq:emission}. Changing variables from \(\log r_{ti}\) to \(r_{ti}\)
contributes the factor \(1/r_{ti}\).
Evaluating the resulting density at the observed return gives the model
likelihood:
\begin{equation}
 \ell_{kti}=
 \frac{1}{r_{ti}\sqrt{2\pi v_{ti}}}
 \exp\!\left[-\frac{(\log r_{ti}-\log f_{kti}+v_{ti}/2)^2}{2v_{ti}}\right].
 \label{eq:likelihood}
\end{equation}
A higher \(\ell_{kti}\) means that the model assigns more density to
returns near the observation. Both models use the same noise scale, 
so only the ratio of the two likelihoods matters below. 
With \(x_{ti}=\log r_{ti}\) and \(d_{ti}=\log f^B_{ti}\),
\begin{equation}
 u_{ti}=\log\frac{\ell_{1ti}}{\ell_{0ti}}
 =\frac{(x_{ti}+v_{ti}/2)^2-(x_{ti}-d_{ti}+v_{ti}/2)^2}{2v_{ti}}
 =\frac{d_{ti}(x_{ti}+v_{ti}/2)-d_{ti}^2/2}{v_{ti}}.
 \label{eq:llr}
\end{equation}
The log likelihood ratio is the difference between the squared errors of the two models, scaled by the variance proxy; it is positive when the AOLMA forecast describes the observation better. 
The working model serves only to score the two forecasts: 
it determines \(w_{ti}\), while the portfolio is chosen by NPM. 
Daily log returns have heavier tails than \eqref{eq:emission}; Section~\ref{ssec:robust} replaces the Gaussian density by a Student-\(t\) density.

\section{Adaptive mixture weights}\label{sec:mixture}

For each asset \(i\), we follow its hidden model label \(Z_{ti}\) through
time. Label \(1\) refers to the AOLMA predictive distribution and label
\(0\) to the neutral predictive distribution in \eqref{eq:emission}.
Each asset has its own mixture proportion and soft counts.
The baseline proportion \(\pi_{ti}\) summarizes accumulated predictive
evidence.

\subsection{Learning the baseline mixture proportion}

Before period \(t\), let \(M_{1ti}\) and \(M_{0ti}\) be the discounted
soft counts assigned to the AOLMA and neutral models for asset \(i\).
Their relative sizes determine the baseline AOLMA proportion:
\begin{equation}
 \pi_{ti}=\frac{M_{1ti}}{M_{1ti}+M_{0ti}}.
 \label{eq:pi}
\end{equation}
Let \(t_0\) denote the first period whose return enters gate calibration.
Before observing that return, initialize
\(M_{1,t_0,i}=M_{0,t_0,i}=1\), giving \(\pi_{t_0,i}=1/2\).
These equal pseudocounts assign equal initial weight to the two models.
The calibration dates are specified in Section~\ref{sec:settings}.

After observing \(r_{ti}\), evaluate the two likelihoods
\(\ell_{1ti}\) and \(\ell_{0ti}\) using Eq. \eqref{eq:likelihood}.
Under the baseline mixture, the posterior probability assigned to model
\(1\) is given by
\begin{equation}
 z^{\mathrm M}_{ti}=
 \frac{\pi_{ti}\ell_{1ti}}
 {\pi_{ti}\ell_{1ti}+(1-\pi_{ti})\ell_{0ti}}
 =\frac{\pi_{ti}e^{u_{ti}}}{\pi_{ti}e^{u_{ti}}+1-\pi_{ti}}.
 \label{eq:responsibility}
\end{equation}
Superscript \(\mathrm M\) identifies the adaptive mixture.
This posterior probability is also called its responsibility for the
observation. 
The observation adds \(z^{\mathrm M}_{ti}\) to the AOLMA
count and \(1-z^{\mathrm M}_{ti}\) to the neutral count.
With forgetting factor \(\rho_M=0.98\), update
\begin{equation}
 \begin{aligned}
 M_{1,t+1,i}&=\rho_M M_{1ti}+z^{\mathrm M}_{ti},\\
 M_{0,t+1,i}&=\rho_M M_{0ti}+1-z^{\mathrm M}_{ti}.
 \end{aligned}
 \label{eq:mixture}
\end{equation}
This uses posterior responsibilities as fractional counts, 
as in the online mixture estimation \cite{cappe2009line}. 
Before period \(t\), a responsibility computed in period \(s<t\) has remaining weight \(\rho_M^{t-1-s}\). 
Each initial pseudo-count has remaining weight
\(\rho_M^{t-t_0}\). 
Thus, \(\pi_{ti}\) adapts over time using discounted
past observations; the total count tends to \(1/(1-\rho_M)=50\), and each observation moves \(\pi_{ti}\) by at most about \(1/50\) in the long run.
The adaptive-mixture gate uses \(w_{ti}=\pi_{ti}\) in Eq. \eqref{eq:forecast}. 
Dynamic model averaging likewise updates model weights using predictive likelihoods and forgetting \cite{raftery2010online}.

\subsection{The mixture proportion estimations}

\begin{prop}\label{prop:pibeta}
Suppose that, given the past, the log return \(x_{ti}\) is Gaussian with mean \(\beta d_{ti}-v_{ti}/2\) and variance \(v_{ti}\). 
Let \(\varepsilon_{ti}=d_{ti}/\sqrt{v_{ti}}\) be the signal-to-noise ratio 
and \(n_{ti}=M_{1ti}+M_{0ti}\). Then, we have
\[
 \E[\pi_{t+1,i}-\pi_{ti}\mid\text{past}]
 =\frac{\pi_{ti}(1-\pi_{ti})(\beta-\pi_{ti})}{\rho_Mn_{ti}+1}\,\varepsilon_{ti}^2+O(\varepsilon_{ti}^4).
\]
To leading order, \(\pi_{ti}\) thus moves towards \(\beta\) (towards \(0\) if \(\beta\leq0\) and towards \(1\) if \(\beta\geq1\)). At \(\pi_{ti}=\beta\), the combined forecast \(\widehat f_{ti}=1+\beta(f^B_{ti}-1)\) equals \(\E[r_{ti}\mid\text{past}]=(f^B_{ti})^\beta\) up to \(O(d_{ti}^2)\).
\end{prop}

The proof is in Appendix~\ref{app:pibeta}. 
The adaptive mixture is therefore an online, 
asset-specific estimate of the share of the forecast that is realized, 
and Eq. \eqref{eq:forecast} rescales each forecast by it: this is the second remedy of Section~\ref{sec:prelim}. 
Because each responsibility lies in \([0,1]\), a single extreme return changes the counts by at most one, and \(\pi_{ti}\) by at most about \(1/50\) in the long run.

\subsection{The combined forecast in the NPM rule}

Substituting the background recurrence \eqref{eq:background} into \eqref{eq:forecast}, \(\widehat f_{ti}=1+w_{ti}(1-\alpha^B_{ti})(y^B_{ti}-1)\), which gives
\begin{equation}
 \widehat f_{ti}
 =\alpha^{\mathrm{eff}}_{ti}
  +(1-\alpha^{\mathrm{eff}}_{ti})y^B_{ti},
 \qquad
 \alpha^{\mathrm{eff}}_{ti}
 =w_{ti}\alpha^B_{ti}+(1-w_{ti}).
 \label{eq:alpha}
\end{equation}
Thus, the gate changes the effective decay factor through its predictive
model weight. 
Since \(y^B_{ti}\) still follows the background recurrence, 
the gate keeps the memory of the moving average and 
shrinks only the size of its signal.

\begin{prop}\label{prop:gatenpm}
Apply the NPM rule \eqref{eq:npm} to the combined forecasts \eqref{eq:forecast}.
\begin{enumerate}[label=(\alph*)]
\item If \(w_{ti}=w>0\) for all \(i\), the solution equals that of NPM on \(\mathbf f^B_t\) with threshold \(c/w\).
\item In general, \(j^*\) maximizes \(w_{ti}(f^B_{ti}-1)\), and asset \(i\) is sold if and only if \(w_{tj^*}(f^B_{tj^*}-1)-w_{ti}(f^B_{ti}-1)>c\).
\item If \(w_{ti}=0\) for all \(i\), NPM does not trade.
\end{enumerate}
\end{prop}

\begin{proof}
(a) Since \(\mathbf b^{\mathsf T}\mathbf 1=1\), the objective equals 
$$1+w[\mathbf b^{\mathsf T}(\mathbf f^B_t-\mathbf 1)-\frac{c}{2w}\|\mathbf b-\mathbf h_t\|_1],
$$ 
and Proposition~\ref{prop:npm} applies with cost \(c/w\). (b) and (c) follow from Proposition~\ref{prop:npm}, 
since \(\widehat f_{tj}-\widehat f_{ti}=w_{tj}(f^B_{tj}-1)-w_{ti}(f^B_{ti}-1)\).
\end{proof}

A common weight \(w\) thus raises the threshold to \(c/w\), whereas asset-specific weights also change which asset looks best: the gate re-ranks the assets by the reliability of their reversal signals.

\section{A payoff-weighted fund of gated experts}\label{sec:fund}

AOLNPM tunes its decay factors to the forecast error with a dataset-specific step size. We instead run NPM experts with fixed decay factors on a grid and weight them by the wealth they earn.

\subsection{Experts}

Let \(\mathcal A=\{0.05,0.10,\ldots,0.95\}\). For each \(\alpha\in\mathcal A\), expert \(\alpha\) runs the recurrence \eqref{eq:background} with the fixed decay factor \(\alpha^B_{ti}=\alpha\), the adaptive mixture of Section~\ref{sec:mixture} with its own soft counts, and the NPM rule on its combined forecasts \(\widehat{\mathbf f}^{(\alpha)}_t\). It keeps its own pre-trade weights \(\mathbf h^{(\alpha)}_t\) and wealth
\begin{equation}
 S^{(\alpha)}_t=S^{(\alpha)}_{t-1}\left(\mathbf b^{(\alpha)\mathsf T}_t\mathbf r_t-\frac c2\|\mathbf b^{(\alpha)}_t-\mathbf h^{(\alpha)}_t\|_1\right),\qquad S^{(\alpha)}_0=1.
 \label{eq:expertwealth}
\end{equation}
The variance proxy \eqref{eq:variance} depends only on returns and is shared by the experts.

\subsection{The fund}

The fund weights the experts by the wealth they have earned so far,
\begin{equation}
 \lambda^{(\alpha)}_t=\frac{S^{(\alpha)}_{t-1}}{\sum_{\alpha'\in\mathcal A}S^{(\alpha')}_{t-1}},\qquad
 \mathbf b_t=\sum_{\alpha\in\mathcal A}\lambda^{(\alpha)}_t\mathbf b^{(\alpha)}_t,
 \label{eq:fund}
\end{equation}
and trades from its own pre-trade weights \(\mathbf h_t\), paying \(\frac c2\|\mathbf b_t-\mathbf h_t\|_1\) as in \eqref{eq:wealth}. 
Trades of different experts in opposite directions cancel. 
The weights \eqref{eq:fund} are the posterior over experts when log wealth plays the role of the log likelihood, as in the universal portfolio over finitely many strategies \cite{cover1991universal,cesa2006prediction,blum1999universal}; they need no learning rate. 
We call the result gated payoff-weighted NPM (G-PWNPM). 
Without the gate (\(w^{(\alpha)}_{ti}=1\)) it is a payoff-weighted NPM (PWNPM).

\begin{prop}\label{prop:fund}
Let \(W^{\mathrm{sep}}_t=|\mathcal A|^{-1}\sum_{\alpha}S^{(\alpha)}_t\) be the wealth of an investor who splits the initial wealth equally among the experts and keeps them in separate accounts. Then
\begin{enumerate}[label=(\roman*)]
\item \(\log W^{\mathrm{sep}}_T\geq\max_{\alpha}\log S^{(\alpha)}_T-\log|\mathcal A|\);
\item in period \(t\) this investor holds \(\mathbf b_t\) of \eqref{eq:fund} in aggregate, with aggregate pre-trade weights \(\sum_\alpha\lambda^{(\alpha)}_t\mathbf h^{(\alpha)}_t\);
\item executing only the net of the accounts' trades costs at most the sum of their separate costs.
\end{enumerate}
\end{prop}

\begin{proof}
(i) A sum of non-negative terms is at least its largest term. \\
(ii) Account \(\alpha\) invests \(|\mathcal A|^{-1}S^{(\alpha)}_{t-1}\mathbf b^{(\alpha)}_t\); after the previous period it held \(|\mathcal A|^{-1}S^{(\alpha)}_{t-1}\mathbf h^{(\alpha)}_t\), since costs are taken in proportion to holdings. Dividing by \(\sum_{\alpha'}|\mathcal A|^{-1}S^{(\alpha')}_{t-1}\) gives both claims.\\ 
(iii) $$\frac{c}{2}\|\sum_\alpha\lambda^{(\alpha)}_t(\mathbf b^{(\alpha)}_t-\mathbf h^{(\alpha)}_t)\|_1\leq\frac{c}{2}\sum_\alpha\lambda^{(\alpha)}_t\|\mathbf b^{(\alpha)}_t-\mathbf h^{(\alpha)}_t\|_1.
$$
\end{proof}

The fund nets the trades of the separate accounts. Its own pre-trade weights differ from the aggregate weights in (ii) only through the experts' different costs in the previous period, so (i) holds for the fund up to terms of order \(c\); in our experiments the netted fund ends above the separate accounts on every dataset.

\subsection{The algorithm}

Algorithm~\ref{alg:gpwnpm} summarizes G-PWNPM. Every decision uses \(r_1,\ldots,r_{t-1}\) only. Each period costs \(O(|\mathcal A|\,n)\) operations, and our Python implementation runs each dataset in under one second.

\begin{algorithm}[htbp]
\caption{Gated payoff-weighted NPM (G-PWNPM)}
\label{alg:gpwnpm}
\begin{algorithmic}[1]
\Require cost rate \(c\); grid \(\mathcal A\); \(\rho_v=\rho_M=0.98\); calibration from period \(t_0\); gate used from period \(t_g\)
\State \(S^{(\alpha)}\gets1\), \(\mathbf h^{(\alpha)}\gets\mathbf 1/n\), \(M^{(\alpha)}_{1i}=M^{(\alpha)}_{0i}\gets1\) for all \(\alpha,i\); \(\mathbf h\gets\mathbf 1/n\), \(W\gets1\)
\For{\(t=1,\ldots,T\)}
  \State \(f^{(\alpha)}_{ti}\gets\) recurrence \eqref{eq:background} with decay \(\alpha\) (forecast \(1\) in the first six periods)
  \State \(w^{(\alpha)}_{ti}\gets M^{(\alpha)}_{1i}/(M^{(\alpha)}_{1i}+M^{(\alpha)}_{0i})\) if \(t\geq t_g\), else \(1\); \(\widehat f^{(\alpha)}_{ti}\gets1+w^{(\alpha)}_{ti}(f^{(\alpha)}_{ti}-1)\)
  \State \(\mathbf b^{(\alpha)}_t\gets\) Proposition~\ref{prop:npm} applied to \(\widehat{\mathbf f}^{(\alpha)}_t\) and \(\mathbf h^{(\alpha)}\) (equal weights in the first six periods)
  \State \(\lambda^{(\alpha)}_t\gets S^{(\alpha)}/\sum_{\alpha'}S^{(\alpha')}\); \(\mathbf b_t\gets\sum_\alpha\lambda^{(\alpha)}_t\mathbf b^{(\alpha)}_t\)
  \State observe \(\mathbf r_t\); \(W\gets W(\mathbf b_t^{\mathsf T}\mathbf r_t-\frac c2\|\mathbf b_t-\mathbf h\|_1)\); \(\mathbf h\gets\mathbf b_t\odot\mathbf r_t/\mathbf b_t^{\mathsf T}\mathbf r_t\)
  \State update \(S^{(\alpha)}\) by \eqref{eq:expertwealth} and \(\mathbf h^{(\alpha)}\gets\mathbf b^{(\alpha)}_t\odot\mathbf r_t/\mathbf b^{(\alpha)\mathsf T}_t\mathbf r_t\)
  \State if \(t\geq t_0\): update \(M^{(\alpha)}_{1i},M^{(\alpha)}_{0i}\) by \eqref{eq:responsibility}--\eqref{eq:mixture} with \(f_{1ti}=f^{(\alpha)}_{ti}\)
  \State update \(v_{ti}\) by \eqref{eq:variance}
\EndFor
\end{algorithmic}
\end{algorithm}

\section{Numerical results}\label{sec:results}

\subsection{Data and settings}\label{sec:settings}

Nine datasets are displayed in Table~\ref{tab:data}. The first four datasets TSE, MSCI, NYSE-N and NYSE-O measured as return periods by assets, respectively, have sizes \(1259\times88\), \(1043\times24\), \(6431\times23\) and \(5651\times36\). They are also the benchmark datasets used in \cite{guo2021adaptive}. 
Their background learning rates are \(5\times10^{-5}\) for TSE,
\(6\times10^{-4}\) for MSCI, \(10^{-5}\) for NYSE-N and
\(2\times10^{-5}\) for NYSE-O, as in \cite{guo2021adaptive}. 
The additional complete panels DJIA, SP500, SZ50, COVID and new-DJIA use the same gate settings and background learning rate \(5\times10^{-5}\). The background factor starts at \(0.5\), and is updated from period 2
and resets to \(0.5\) on crossing a boundary. 
Furthermore, all NPM strategies use forecasts \(1\) and equal-weight portfolios for the first six periods.

\begin{table}[htbp]
\centering
\small
\begin{tabular}{llrrr}
\toprule
Dataset & Description & Assets & Periods & \(\gamma\)\\
\midrule
TSE & Toronto Stock Exchange stocks, 1994--1998 & 88 & 1259 & $5\times10^{-5}$ \\
MSCI & MSCI World constituent indices, 2006--2010 & 24 & 1043 & $6\times10^{-4}$ \\
NYSE-N & New York Stock Exchange stocks, 1985--2010 & 23 & 6431 & $1\times10^{-5}$ \\
NYSE-O & New York Stock Exchange stocks, 1962--1984 & 36 & 5651 & $2\times10^{-5}$ \\
\midrule
DJIA & Dow Jones Industrial Average stocks, 2001--2003 & 30 & 507 & $5\times10^{-5}$ \\
SP500 & S\&P 500 stocks, 1998--2003 & 25 & 1276 & $5\times10^{-5}$ \\
SZ50 & SSE 50 stocks, 2014--2018 & 26 & 1219 & $5\times10^{-5}$ \\
COVID & Dow Jones Industrial Average stocks, 2019--2023 & 30 & 1189 & $5\times10^{-5}$ \\
new-DJIA & Dow Jones Industrial Average stocks, 2008--2016 & 30 & 2115 & $5\times10^{-5}$ \\
\bottomrule
\end{tabular}
\caption{Datasets. Periods count price relatives; \(\gamma\) is the background learning rate of AOLNPM.}
\label{tab:data}
\end{table}

Periods are numbered from 1. 
All G-PWNPM experts in Algorithm~\ref{alg:gpwnpm} adopt the same timing. 
Gate calibration begins with the return observed in period 21, so $t_0=21$. Before this return period, the two soft counts for each asset and expert are initialized to 1. 
The learned gate weights are first used to adjust forecasts for trading in period 102, so $t_g=102$. 
The transaction cost rate is $c=0.001$ unless otherwise stated.

\subsection{Comparisons and evaluations}


We compare thirteen baselines. 
The benchmarks are the market (buys and holds the uniform portfolio), the uniform constant rebalanced portfolio (UCRP) and the best constant rebalanced portfolio in hindsight (BCRP) \cite{cover1991universal}; BCRP uses future prices and serves only as a reference. 
The follow-the-winner strategies are EG \cite{helmbold1998line} and ONS \cite{agarwal2006algorithms}. The follow-the-loser strategies are Anticor \cite{borodin2004can}, PAMR \cite{li2012pamr}, OLMAR-1 and OLMAR-2 \cite{li2015moving}, and RMR \cite{huang2016robust}. 
The cost-aware strategies are TCO1 and TCO2 \cite{li2018transaction} and AOLNPM \cite{guo2021adaptive}. Every strategy pays the costs in \eqref{eq:wealth}. Appendix~\ref{app:baselines} gives the parameters and compares our results with \cite[Table~2]{guo2021adaptive}.

Three strategies are involved as our contribution for comparison. Applying the gate of Section~\ref{sec:mixture} to the forecasts of AOLNPM directly gives AOLNPM-AM. PWNPM is the fund of Section~\ref{sec:fund} without the gate. 
Finally, our proposed method is G-PWNPM with the grid \(\mathcal A\) of \(19\) decay factors.

For two strategies with daily log growth \(g_t=\log(W_t/W_{t-1})\) and \(g'_t\), the ratio of their terminal wealth is \(\exp\sum_t(g_t-g'_t)\). We test whether the mean of \(g_t-g'_t\) is positive with the \(t\)-statistic based on the Newey--West standard error with 20 lags \cite{newey1987simple}. Across the nine datasets we report the geometric mean of the ratios, the number of datasets with ratio above \(1\) and the pooled statistic \(z=\sum_d t_d/3\). If the datasets were independent, \(z\) would be approximately standard normal under the null hypothesis; some datasets overlap in time, so \(z\) is indicative. We also report the average one-way turnover \(\|\mathbf b_t-\mathbf h_t\|_1/2\) after the warm-up.

\subsection{Terminal wealth}

\begin{figure}[htbp]
\centering
\includegraphics[width=\linewidth]{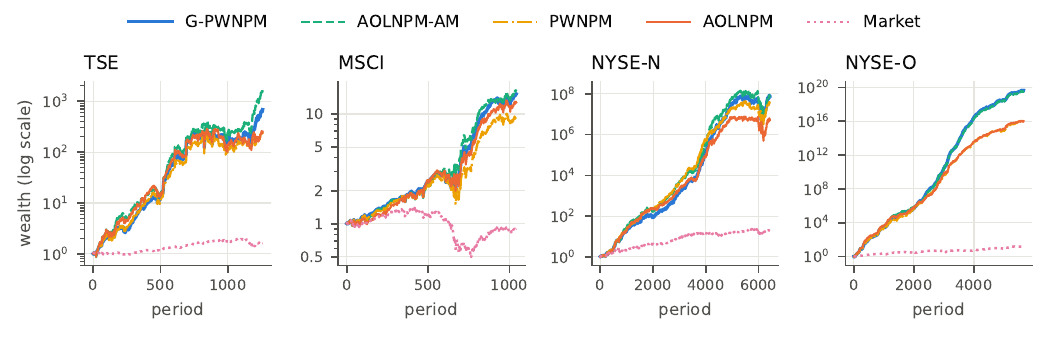}
\caption{Wealth paths on the benchmark datasets from initial wealth \(1\) at \(c=0.001\), with logarithmic vertical
axes.}
\label{fig:wealth}
\end{figure}

Tables~\ref{tab:wealth} and~\ref{tab:wealth2} report terminal wealth from initial wealth \(1\), and Figure~\ref{fig:wealth} shows the wealth paths on the benchmark datasets. 
On the four benchmark datasets, G-PWNPM ends above all thirteen baselines, with \(2.7\), \(1.2\), \(15\) and \(4{,}567\) times the terminal wealth of AOLNPM on TSE, MSCI, NYSE-N and NYSE-O. 
The gate alone also improves on AOLNPM there, and AOLNPM-AM has the highest terminal wealth on TSE and MSCI.

\begin{table}[htbp]
\centering
\small
\begin{tabular}{lrrrr}
\toprule
Strategy & TSE & MSCI & NYSE-N & NYSE-O\\
\midrule
Market & 1.613 & 0.906 & 18.06 & 14.50 \\
UCRP & 1.581 & 0.923 & 30.35 & 26.20 \\
BCRP & 6.722 & 1.503 & 115.7 & 235.0 \\
\addlinespace[3pt]
EG & 1.580 & 0.922 & 29.88 & 26.25 \\
ONS & 1.524 & 0.826 & 18.95 & 94.00 \\
\addlinespace[3pt]
Anticor & 33.48 & 2.448 & 61{,}449 & $1.59\!\times\!10^{7}$ \\
PAMR & 101.4 & 6.045 & 5{,}689 & $4.43\!\times\!10^{13}$ \\
OLMAR-1 & 27.06 & 7.334 & $7.69\!\times\!10^{6}$ & $1.77\!\times\!10^{15}$ \\
OLMAR-2 & 262.9 & 9.992 & $4.10\!\times\!10^{6}$ & $1.21\!\times\!10^{16}$ \\
RMR & 81.20 & 8.060 & $5.24\!\times\!10^{6}$ & $3.68\!\times\!10^{15}$ \\
\addlinespace[3pt]
TCO1 & 60.90 & 6.004 & $4.96\!\times\!10^{5}$ & $5.98\!\times\!10^{12}$ \\
TCO2 & 67.89 & 3.982 & $3.64\!\times\!10^{6}$ & $9.08\!\times\!10^{11}$ \\
AOLNPM & 251.4 & 12.73 & $5.23\!\times\!10^{6}$ & $1.03\!\times\!10^{16}$ \\
\addlinespace[3pt]
PWNPM & 251.0 & 9.498 & $3.36\!\times\!10^{7}$ & $1.07\!\times\!10^{16}$ \\
AOLNPM-AM & {\boldmath\bfseries 1{,}642} & {\boldmath\bfseries 16.47} & $7.23\!\times\!10^{7}$ & $3.64\!\times\!10^{19}$ \\
G-PWNPM & 683.0 & 15.21 & {\boldmath\bfseries $7.90\!\times\!10^{7}$} & {\boldmath\bfseries $4.69\!\times\!10^{19}$} \\
\bottomrule
\end{tabular}
\caption{Terminal wealth on the benchmark datasets, \(W_0=1\) and \(c=0.001\).
Bold indicates the highest terminal wealth of an online strategy in each column; BCRP uses future prices.}
\label{tab:wealth}
\end{table}

\begin{table}[htbp]
\centering
\small
\begin{tabular}{lrrrrr}
\toprule
Strategy & DJIA & SP500 & SZ50 & COVID & new-DJIA\\
\midrule
Market & 0.764 & 1.342 & 2.037 & 1.418 & 2.192 \\
UCRP & 0.810 & 1.633 & 2.137 & 1.427 & 2.368 \\
BCRP & 1.235 & 4.033 & 6.230 & 4.009 & 5.484 \\
\addlinespace[3pt]
EG & 0.807 & 1.618 & 2.131 & 1.427 & 2.360 \\
ONS & 1.481 & 3.162 & {\boldmath\bfseries 2.713} & {\boldmath\bfseries 1.471} & {\boldmath\bfseries 4.762} \\
\addlinespace[3pt]
Anticor & 1.506 & 4.411 & 1.913 & 1.198 & 2.667 \\
PAMR & 0.451 & 1.805 & 0.101 & 0.081 & 0.101 \\
OLMAR-1 & 1.593 & 7.416 & 0.449 & 0.536 & 0.893 \\
OLMAR-2 & 0.790 & 3.557 & 0.254 & 0.235 & 0.274 \\
RMR & 1.708 & 4.026 & 0.331 & 0.365 & 1.013 \\
\addlinespace[3pt]
TCO1 & 1.166 & 2.725 & 0.475 & 0.360 & 1.231 \\
TCO2 & {\boldmath\bfseries 1.741} & {\boldmath\bfseries 8.184} & 1.005 & 0.741 & 2.502 \\
AOLNPM & 0.901 & 2.671 & 0.257 & 0.259 & 0.250 \\
\addlinespace[3pt]
PWNPM & 1.109 & 3.750 & 0.292 & 0.356 & 0.592 \\
AOLNPM-AM & 0.994 & 3.488 & 0.377 & 0.216 & 0.893 \\
G-PWNPM & 1.058 & 3.426 & 0.770 & 0.396 & 1.437 \\
\bottomrule
\end{tabular}
\caption{Terminal wealth on the additional datasets, \(W_0=1\) and \(c=0.001\).
Bold indicates the highest terminal wealth of an online strategy in each column.}
\label{tab:wealth2}
\end{table}

On the five additional datasets the AOLMA forecasts carry little information 
(Table~\ref{tab:calib}), and every follow-the-loser strategy except Anticor loses money on SZ50 and COVID. 
G-PWNPM still ends above AOLNPM on all five, by factors between \(1.17\) (DJIA) and \(5.75\) (new-DJIA), but it ranks in the middle of the baselines. 
TCO2 ends highest on DJIA and SP500. 
On SZ50, COVID and new-DJIA, where the market rises, the benchmarks, the follow-the-winner strategies, Anticor and TCO2 all end above G-PWNPM.

\subsection{Significance and turnover}

Table~\ref{tab:sig} compares G-PWNPM with AOLNPM on each dataset. 
G-PWNPM ends above AOLNPM on all nine datasets, with geometric mean ratio \(5.92\) and pooled \(z=4.68\). 
On a single dataset the daily differences are noisy; the difference is significant at the \(5\%\) level on 
NYSE-O, SZ50 and new-DJIA. 
Against the best baseline of each benchmark dataset (OLMAR-2 on TSE and NYSE-O, AOLNPM on MSCI and OLMAR-1 on NYSE-N), 
the ratios are \(2.60\), \(1.20\), \(10.3\) and \(3{,}878\), with \(t=0.89\), \(0.66\), \(1.17\) and \(3.80\).

\begin{table}[htbp]
\centering
\small
\setlength{\tabcolsep}{3pt}
\begin{tabular}{lrrrrrrrrr}
\toprule
& TSE & MSCI & NYSE-N & NYSE-O & DJIA & SP500 & SZ50 & COVID & new-DJIA\\
\midrule
Wealth ratio & 2.72 & 1.20 & 15.1 & 4{,}567 & 1.17 & 1.28 & 2.99 & 1.53 & 5.75 \\
$t$-statistic & $0.92$ & $0.66$ & $1.40$ & $3.91$ & $0.57$ & $0.33$ & $2.14$ & $1.09$ & $3.01$ \\
\addlinespace[2pt]
Turnover, AOLNPM & 0.78 & 0.78 & 0.72 & 0.78 & 0.74 & 0.75 & 0.69 & 0.70 & 0.70 \\
Turnover, G-PWNPM & 0.66 & 0.49 & 0.40 & 0.48 & 0.53 & 0.52 & 0.31 & 0.46 & 0.32 \\
\bottomrule
\end{tabular}
\caption{G-PWNPM against AOLNPM at \(c=0.001\). Wealth ratio: terminal wealth of G-PWNPM over that of AOLNPM; \(t\)-statistic: Newey--West \(t\)-statistic of the difference in daily log growth; turnover: average one-way turnover per period.}
\label{tab:sig}
\end{table}

G-PWNPM trades less than AOLNPM on every dataset, with average 
one-way turnover \(15\)--\(55\%\) below that of AOLNPM. 
Turnover nevertheless remains high, \(0.31\)--\(0.66\) of the portfolio per period against \(0.69\)--\(0.78\) for AOLNPM, because NPM moves all sold holdings into a single asset. 
G-PWNPM also has a higher annualized Sharpe ratio than AOLNPM on all nine datasets and a smaller maximum drawdown on eight.

\subsection{Transaction costs and ablation}\label{ssec:costs}

Table~\ref{tab:costs} varies the cost rate 
from \(0\) to \(0.7\%\), which covers the range \(0.1\)--\(0.7\%\) of \cite{guo2021adaptive}. 
The advantage of G-PWNPM over AOLNPM grows with the cost rate, from \(3.02\) without costs to \(5.92\) at \(c=0.1\%\) and \(131\) at \(c=0.7\%\) (geometric means), and G-PWNPM ends above AOLNPM on every dataset for \(c\geq0.1\%\). 
Without costs, NPM holds only the asset with the highest forecast, so the gain of AOLNPM-AM at \(c=0\) (\(2.97\)) comes from the re-ranking of Proposition~\ref{prop:gatenpm}(b) alone. 
The fund alone also gains on AOLNPM as the cost rate grows (PWNPM, from \(1.11\) to \(14.16\)). 
Among the baselines, OLMAR-2 and RMR fall far behind AOLNPM at high cost rates. TCO2, which also controls costs, overtakes 
AOLNPM from \(c=0.2\%\) but stays below G-PWNPM in geometric mean at every cost rate.

\begin{table}[htbp]
\centering
\small
\begin{tabular}{lrrrrrrr}
\toprule
& \multicolumn{7}{c}{Cost rate \(c\)}\\
\cmidrule(lr){2-8}
Strategy & \(0\) & \(0.05\%\) & \(0.1\%\) & \(0.2\%\) & \(0.3\%\) & \(0.5\%\) & \(0.7\%\)\\
\midrule
OLMAR-2 & 1.09 (7) & 1.04 (7) & 0.98 (4) & 0.81 (3) & 0.57 (2) & 0.16 (0) & 0.03 (0) \\
RMR & 0.91 (4) & 0.98 (6) & 1.05 (6) & 1.11 (5) & 1.00 (5) & 0.47 (2) & 0.13 (2) \\
TCO2 & 0.21 (5) & 0.34 (5) & 0.53 (5) & 1.26 (6) & 2.62 (6) & 7.03 (8) & 12.40 (8) \\
PWNPM & 1.11 (5) & 1.28 (6) & 1.47 (7) & 2.21 (7) & 3.19 (8) & 7.21 (9) & 14.16 (9) \\
AOLNPM-AM & 2.97 (7) & 3.98 (8) & 5.16 (8) & 12.93 (8) & 25.63 (8) & 47.84 (9) & 89.52 (9) \\
G-PWNPM & 3.02 (8) & 4.41 (8) & 5.92 (9) & 14.80 (9) & 30.70 (9) & 71.41 (9) & 131 (9) \\
\bottomrule
\end{tabular}
\caption{Terminal wealth relative to AOLNPM at the same cost rate: geometric mean over the nine datasets and, in parentheses, the number of datasets with ratio above \(1\).}
\label{tab:costs}
\end{table}

The rows PWNPM and AOLNPM-AM of Table~\ref{tab:costs} separate the fund and the gate. 
The gate supplies most of the gain: at \(c=0.1\%\), applied to the forecasts of AOLNPM alone (AOLNPM-AM), it multiplies terminal wealth by \(5.16\) (geometric mean, pooled \(z=3.53\)), against \(1.47\) for the fund alone (PWNPM, \(z=1.72\)). Combining the two, G-PWNPM reaches \(5.92\); it is the only strategy above AOLNPM on all nine datasets and has the largest pooled \(z\). 
Against PWNPM, the gate multiplies terminal wealth by \(4.03\) (\(z=4.23\)); against AOLNPM-AM, the fund adds less on average (\(1.15\), \(z=1.16\)). The fund mainly adds consistency: on the five additional datasets, the geometric mean rises from \(1.44\) for AOLNPM-AM to \(2.09\), whereas on the benchmark datasets AOLNPM-AM is higher (\(25.36\) against \(21.75\)). Without NYSE-O, the dataset with the largest gain, G-PWNPM still gains \(2.58\). The fund ends within a factor \(1.85\)--\(5.81\) of its best expert in hindsight, well inside the factor \(|\mathcal A|=19\) of Proposition~\ref{prop:fund}(i).

Proposition~\ref{prop:gatenpm} separates two effects of the gate: a common weight raises the NPM threshold, and asset-specific weights re-rank the assets. 
Replacing the weights of AOLNPM-AM from period 102 isolates them. A common constant weight $w$ is equivalent to using a threshold 
$c/w$ on the original forecasts. 
With $w=1/8$ on all nine datasets, the geometric mean terminal wealth ratio relative to AOLNPM is $2.15$. 
Selecting the best-performing weight in hindsight for each dataset from $\{1/10, 1/8, 1/6, 1/5, 1/4, 1/3, 1/2, 1\}$ raises this ratio to $2.33$. 
Keeping the cross-sectional variation of \(\pi_{ti}\) at the full-sample mean level recovers most of the gain of the adaptive mixture (\(4.42\) against \(5.16\)), whereas giving each asset the proportions of another asset loses wealth (\(0.13\)). 
The re-ranking of the assets by the reliability of their signals is therefore the main effect.

\subsection{Robustness}\label{ssec:robust}

Changing one setting of G-PWNPM at a time, namely the count memory \(\rho_M\in\{0.95,0.97,0.99,0.995\}\), the variance memory \(\rho_v\in\{0.94,0.97,0.99\}\), the start of the gate (trading from period 52 or 202, or counts from \(r_{51}\)) or the grid (\(9\), \(13\) or \(39\) decay factors), keeps the geometric mean ratio over AOLNPM between \(5.44\) and \(6.48\), and G-PWNPM stays above AOLNPM on at least seven datasets. A Student-\(t\) likelihood with \(\nu\in\{3,4\}\) degrees of freedom and variance \(v_{ti}\) raises the geometric mean ratio to \(7.00\)--\(7.72\), but G-PWNPM then falls slightly below AOLNPM on SP500, and for \(\nu=3\) also on MSCI; we keep the Gaussian likelihood of Section~\ref{sec:models}. G-PWNPM concentrates its holdings: after period 101, its average weight on its most-held stock ranges from \(8\%\) (new-DJIA) to \(42\%\) (NYSE-O). When that stock is removed from each dataset and both strategies are rerun, G-PWNPM still ends above AOLNPM on eight datasets, with geometric mean ratio \(2.88\); on NYSE-O the ratio falls from \(4{,}567\) to \(16.8\), and on TSE it falls to \(0.60\). The gains thus do not rest on a single stock, although a large part of the gain on NYSE-O does.

\section{Conclusions}\label{sec:conclusion}

NPM trades whenever a forecast gap exceeds the cost, so its control of costs is only as good as the calibration of its forecasts, and AOLMA forecasts are far from calibrated. 
For each asset, we learn online the proportion of an adaptive mixture of the AOLMA forecast and a neutral forecast from their predictive likelihoods, and shrink the forecast by this proportion before NPM. 
The proportion estimates the share of the forecast that is realized. 
For NPM the gate acts as an asset-specific threshold, and its re-ranking of the assets is the main source of the gain. 
A payoff-weighted fund of gated experts with fixed decay factors replaces the dataset-specific step size of AOLNPM. On nine datasets, G-PWNPM ends above AOLNPM on all nine, with lower turnover and, on most datasets, smaller drawdowns, and its advantage grows with the cost rate.

The gate learns slowly, over several hundred periods (Appendix~\ref{app:pibeta}), and on some datasets where AOLMA has little predictive power, G-PWNPM reduces the losses of NPM without turning them into gains. 
Faster learning of the proportions, for example by pooling information across assets, and funds that also hold 
follow-the-winner experts are natural extensions.

\appendix

\section{Proof of Proposition~\ref{prop:pibeta}}\label{app:pibeta}

\begin{proof}
Drop the indices \(t\) and \(i\), and write \(x+v/2=\beta d+\sqrt v\,\xi\) with \(\xi\sim N(0,1)\). 
By Eq. \eqref{eq:llr},
\[
 u=\frac{d(\beta d+\sqrt v\,\xi)-d^2/2}{v}=\varepsilon\xi+\bigl(\beta-\tfrac12\bigr)\varepsilon^2,
\]
so \(\E u=(\beta-\frac12)\varepsilon^2\), \(\E u^2=\varepsilon^2+O(\varepsilon^4)\), \(\E u^3=O(\varepsilon^4)\) and \(\E u^4=O(\varepsilon^4)\). 
By \eqref{eq:responsibility}, \(z=g(u)\) with 
\(g(u)=\pi e^u/(\pi e^u+1-\pi)\), a logistic function of 
\(u+\log\frac{\pi}{1-\pi}\) with bounded derivatives, 
\(g(0)=\pi\), \(g'(0)=\pi(1-\pi)\) and \(g''(0)=\pi(1-\pi)(1-2\pi)\). 
Taylor's theorem with a fourth-order remainder gives
\[
 \E z-\pi=g'(0)\E u+\tfrac12g''(0)\E u^2+\tfrac16g'''(0)\E u^3+O(\E u^4)=\pi(1-\pi)(\beta-\pi)\varepsilon^2+O(\varepsilon^4).
\]
By \eqref{eq:mixture}, 
$\pi_{t+1}=(\rho_Mn \pi+z)/(\rho_Mn+1)$, so 
$\pi_{t+1}-\pi=(z-\pi)/(\rho_Mn+1)$. 
Since \(n\) is determined by the past, the expectation in Proposition~\ref{prop:pibeta} follows. 
The leading term is positive for \(\pi<\beta\) and negative for \(\pi>\beta\); it is negative on \((0,1)\) if \(\beta\leq0\) and positive if \(\beta\geq1\). 
Finally, 
$\E[r\mid\text{past}]=\E e^x=e^{\beta d}=(f^B)^\beta$, and 
$1+\beta(e^d-1)-e^{\beta d}=\frac12\beta(1-\beta)d^2+O(d^3)$.
\end{proof}

Since \(n_{ti}\to1/(1-\rho_M)=50\), the leading term is about \(\pi_{ti}(1-\pi_{ti})(\beta-\pi_{ti})\varepsilon_{ti}^2/50\) per period. For the AOLNPM forecasts, the average of \(\varepsilon_{ti}^2\) is \(0.21\)--\(0.32\) on our datasets, so \(\pi_{ti}\) approaches \(\beta\) over several hundred periods.

\section{Baselines and comparison with \cite{guo2021adaptive}}\label{app:baselines}

All baselines use the default parameters of \cite{li2018online}, as in \cite{guo2021adaptive}, and pay costs as in \eqref{eq:wealth}; parameters are given in the notation of the cited papers. 
The market holds the uniform buy-and-hold portfolio and UCRP rebalances to \(\mathbf 1/n\) every period. 
BCRP is the log-optimal constant rebalanced portfolio without costs, computed in hindsight by Cover's multiplicative updates; it then pays the costs of rebalancing to this portfolio. 
EG uses \(\eta=0.05\), and ONS uses \(\eta=0\), \(\beta=1\) and \(\delta=1/8\). Anticor is the buy-and-hold combination of Anticor experts with windows \(2,\ldots,30\); we weight the experts by their wealth after costs and trade only the net, as in \eqref{eq:fund}. 
PAMR uses \(\varepsilon=0.5\). 
OLMAR-1 uses a 5-period simple moving average and OLMAR-2 an exponential moving average with \(\alpha=0.5\), both with \(\varepsilon=10\). 
RMR uses the \(L_1\)-median of the last 5 prices, computed by Weiszfeld's algorithm, with \(\varepsilon=5\). TCO1 (reversal of the last price relative) and TCO2 (5-period moving-average reversal) use \(\eta=10\) and \(\lambda=10c\) \cite{li2018transaction}.

Table~\ref{tab:check} compares our terminal wealth with \cite[Table~2]{guo2021adaptive}. For the benchmarks and the follow-the-winner strategies the two agree to within \(0.1\%\). For PAMR, OLMAR-1, OLMAR-2 and RMR they differ by at most \(1.5\%\) of log wealth. Anticor differs more, which we attribute to the combination of its experts. Our AOLNPM follows \cite[Section~3]{guo2021adaptive} with the step sizes of Table~\ref{tab:data}, but ends below the published values on all four datasets; we have not found the source of the difference. 
All NPM strategies in this paper share our implementation of AOLMA and NPM, 
so the comparisons with AOLNPM in Section~\ref{sec:results} are made under the same conventions. 
G-PWNPM also ends above the published AOLNPM values on all four datasets: \(683.0\) against \(529.6\) on TSE, \(15.21\) against \(14.94\) on MSCI, \(7.90\times10^7\) against \(7.73\times10^6\) on NYSE-N and \(4.69\times10^{19}\) against \(3.45\times10^{16}\) on NYSE-O.

\begin{table}[H]
\centering
\small
\setlength{\tabcolsep}{3.5pt}
\begin{tabular}{lrrrrrrrr}
\toprule
& \multicolumn{2}{c}{MSCI} & \multicolumn{2}{c}{NYSE-O} & \multicolumn{2}{c}{NYSE-N} & \multicolumn{2}{c}{TSE}\\
\cmidrule(lr){2-3}\cmidrule(lr){4-5}\cmidrule(lr){6-7}\cmidrule(lr){8-9}
Strategy & Ours & \cite{guo2021adaptive} & Ours & \cite{guo2021adaptive} & Ours & \cite{guo2021adaptive} & Ours & \cite{guo2021adaptive}\\
\midrule
Market & 0.906 & 0.906 & 14.50 & 14.50 & 18.06 & 18.06 & 1.613 & 1.613 \\
UCRP & 0.923 & 0.922 & 26.20 & 26.19 & 30.35 & 30.33 & 1.581 & 1.580 \\
BCRP & 1.503 & 1.503 & 235.0 & 235.1 & 115.7 & 115.7 & 6.722 & 6.726 \\
EG & 0.922 & 0.922 & 26.25 & 26.24 & 29.88 & 29.87 & 1.580 & 1.579 \\
ONS & 0.826 & 0.826 & 94.00 & 93.94 & 18.95 & 18.94 & 1.524 & 1.524 \\
Anticor & 2.448 & 3.000 & $1.59\!\times\!10^{7}$ & $4.92\!\times\!10^{6}$ & 61{,}449 & 43{,}957 & 33.48 & 21.23 \\
PAMR & 6.045 & 6.030 & $4.43\!\times\!10^{13}$ & $4.31\!\times\!10^{13}$ & 5{,}689 & 5{,}633 & 101.4 & 100.8 \\
OLMAR-1 & 7.334 & 7.407 & $1.77\!\times\!10^{15}$ & $1.59\!\times\!10^{15}$ & $7.69\!\times\!10^{6}$ & $7.51\!\times\!10^{6}$ & 27.06 & 26.29 \\
OLMAR-2 & 9.992 & 10.03 & $1.21\!\times\!10^{16}$ & $1.17\!\times\!10^{16}$ & $4.10\!\times\!10^{6}$ & $3.97\!\times\!10^{6}$ & 262.9 & 274.5 \\
RMR & 8.060 & 8.123 & $3.68\!\times\!10^{15}$ & $3.48\!\times\!10^{15}$ & $5.24\!\times\!10^{6}$ & $4.18\!\times\!10^{6}$ & 81.20 & 81.86 \\
AOLNPM & 12.73 & 14.94 & $1.03\!\times\!10^{16}$ & $3.45\!\times\!10^{16}$ & $5.23\!\times\!10^{6}$ & $7.73\!\times\!10^{6}$ & 251.4 & 529.6 \\
\bottomrule
\end{tabular}
\caption{Terminal wealth at \(c=0.001\): our implementations and \cite[Table~2]{guo2021adaptive}.}
\label{tab:check}
\end{table}

\noindent
{\bf Acknowledgment:} This research work was supported by 
Hong Kong Research Grants Council under Grant number 17309522.

\bibliographystyle{amsplain-full}
\bibliography{references}

\end{document}